\documentclass[12pt]{article}  

\usepackage{graphicx,color,epsf} 

\usepackage{amsmath,latexsym}

\begin{document}

\begin{center}
{\LARGE \bf Blackbody heat capacity at constant pressure
}
\\ 
\vspace{1cm} 
{\large E. S. Moreira Jr.}
\footnote{E-mail: moreira@unifei.edu.br}   
\\ 
\vspace{0.3cm} 
{\em Instituto de Matem\'{a}tica e Computa\c{c}\~{a}o,}  
{\em Universidade Federal de Itajub\'{a},}   \\
{\em Itajub\'a, Minas Gerais 37500-903, Brazil}

\vspace{0.3cm}
{\large February, 2024}
\end{center}
\vspace{0.5cm}


\abstract{ 
At first glance, the title of this work seems to be improper.
And the reason is well known. Since blackbody pressure depends only on temperature,
one cannot take the derivative of the thermodynamic quantities with respect to one of them, 
keeping the other constant. That is, the heat capacity at constant pressure, $C_{P}$,
as well as, the coefficient of thermal expansion, $\alpha$, and the isothermal compressibility, 
$\kappa_{T}$, are ill-defined quantities. This work will show that when the perfect conducting nature of the walls of a blackbody cavity is taken into account, $C_{P}$, $\alpha$ and $\kappa_{T}$ are in fact well defined, and they are related by the usual thermodynamic relations, as expected.
Two geometries will be considered, namely, a spherical shell and a cubic box.
It will be shown that $C_{P}$, $\alpha$ and $\kappa_{T}$ depend very much on the geometry of the cavity. Issues regarding thermodynamic stability will be addressed,
revealing that they also depend on the cavity's geometry. It is argued that these findings
may be amenable to experimental verification.
}


\section{Introduction}
\label{introduction}
Blackbody radiation is present in most branches of physics. Therefore, a clear understanding of its thermodynamics is vital for the comprehension of many phenomena.
There are many ways to obtain the Helmholtz free energy $F$ of the electromagnetic radiation, at temperature $T$, in an arbitrary large cavity of volume $V$. The most familiar ways are those
in statistical mechanics textbooks (see, e.g., \cite{pat72,cal85,hua87}), which lead to
\footnote{Fundamental constants are the usual ones, and $\beta:=1/kT$.}
\begin{equation}
F=-\frac{\pi^{2}}{45}\frac{(kT)^{4}}{(\hbar c)^{3}} V,
\label{f0}
\end{equation}
yielding the entropy $S=-(\partial_{T}F)_{V}$ and the pressure $P=-(\partial_{V}F)_{T}$,
\begin{equation}
S=\frac{4\pi^{2}}{45}\left(\frac{kT}{\hbar c}\right)^{3}k V, \hspace{2.0cm} 
P=\frac{\pi^{2}}{45}\frac{(kT)^{4}}{(\hbar c)^{3}},
\label{sp0}
\end{equation}
as well as the internal energy $U=\partial_{\beta}(\beta F)$ and the
blackbody heat capacity at constant volume $C=(\partial_{T}U)_{V}$,
\begin{equation}
U=\frac{\pi^{2}}{15}\frac{(kT)^{4}}{(\hbar c)^{3}} V, \hspace{2.0cm} 
C=\frac{4\pi^{2}}{15}k\left(\frac{kT}{\hbar c}\right)^{3}V.
\label{uc0}
\end{equation}
The familiar thermodynamic relations $PV=U/3$ and $C=3S$ are easily verified.

Instead of using $F$ to obtain the blackbody thermodynamics, one could use the 
Gibbs thermodynamic potential $G=F+PV$  and the equations
\begin{equation}
S=-\left(\frac{\partial G}{\partial T}\right)_{P}, \hspace{2.0cm} 
V=\left(\frac{\partial G}{\partial P}\right)_{T}.
\label{sv0}
\end{equation}
However, $G$ vanishes identically, and therefore equation (\ref{sv0}) is not available here.

Other issues arise in the calculation of the heat capacity at constant pressure
\begin{equation}
C_{P}=\left(\frac{\partial H}{\partial T}\right)_{P},
\label{c0}
\end{equation}
where $H=U+PV$ is the enthalpy,
or in the calculations of the coefficient of thermal expansion $\alpha$
and the isothermal compressibility $\kappa_{T}$, which are given by the definitions
\begin{equation}
\alpha:=\frac{1}{V}\left(\frac{\partial V}{\partial T}\right)_{P}, \hspace{2.0cm} 
\kappa_{T}:=-\frac{1}{V}\left(\frac{\partial V}{\partial P}\right)_{T}.
\label{akapa}
\end{equation}
The trouble is that, in blackbody thermodynamics, $P$ and $T$ are not independent parameters [cf. equation (\ref{sp0})],
and thus the quantities in equations (\ref{c0}) and (\ref{akapa}) are ill-defined.
As mentioned in \cite{kel81}, one may try to come to terms with these difficulties by
arguing that  $C_{P}$, $\alpha$ and $\kappa_{T}$ are, in fact,  infinite. Nevertheless, this argument
does not seem to be satisfactory, and there remain other issues \cite{rod89}.

In fact, equation (\ref{f0}) should be rewritten as
\begin{equation}
F=-\frac{\pi^{2}}{45}\frac{(kT)^{4}}{(\hbar c)^{3}} V+\cdots,
\label{f1}
\end{equation}
where 
dots denote vacuum polarization corrections due to the wall's cavity
(see, e.g., \cite{dav82}).  
Usually, such corrections are omitted
in calculations in statistical mechanics, since they are small
at the regime of high temperatures and/or large cavities, i.e., when 
\begin{equation}
\frac{kTV^{1/3}}{\hbar c}\gg 1.
\label{ht}
\end{equation}
However, they do carry the key ingredient which makes 
$C_{P}$, $\alpha$ and $\kappa_{T}$ well defined thermodynamic quantities in the context of 
blackbody radiation, as this work intends to show. 

At this point, it should be remarked that, at the opposite regime 
of low temperatures and/or small cavities (i.e., when $kTV^{1/3}\ll \hbar c$),
equation (\ref{f1}) gives place to $F=U_{0}+\cdots$ where the temperature independent term $U_{0}$
is the vacuum energy in the cavity \cite{dav82}, and now the dots denote thermal corrections.
Thus, the Casimir energy
\footnote{For reviews on the Casimir effect, see \cite{mil01,bor01}.}
$U_{0}$ \cite{cas48,boy68} and equation (\ref{f1}) are, in fact,
two facets of the same phenomenon
\cite{lif56,meh67,bro69,cas70,bal78,sch78,dow78,amb83}.

The text will be organized as follows.
In sections \ref{ss} and \ref{cb}, respectively, the corrections in
equation (\ref{f1}) will be used to calculate well defined $C_{P}$, $\alpha$ and $\kappa_{T}$
for the blackbody radiation in a perfect conducting spherical shell, and in a 
perfect conducting cubic box. In section \ref{ts}, the stability of blackbody radiation
will be determined in each of these cavities. 
In section \ref{fr}, a summary and further analysis concerned the results will be given.
An appendix outlines the derivation of the corrections in  equation (\ref{f1}).

\section{Spherical shell}
\label{ss}
For a perfect conducting spherical shell
carrying hot electromagnetic radiation, equation (\ref{f1}) becomes \cite{bal78}
(see appendix \ref{A})
\begin{equation}
F=-\frac{\pi^{2}}{45}\frac{(kT)^{4}}{(\hbar c)^{3}} V
-\frac{kT}{4}\left[
{\rm ln}\left(\frac{kTV^{1/3}}{\hbar c}\right)
+
0.291
\right],	
\label{fss}
\end{equation}
where smaller  corrections have been neglected.
Accordingly, equation (\ref{sp0}) is replaced by
\begin{eqnarray}
\hspace{-1.0cm} S=\frac{4\pi^{2}}{45}\left(\frac{kT}{\hbar c}\right)^{3}k V
+\frac{k}{4}\left[
{\rm ln}\left(\frac{kTV^{1/3}}{\hbar c}\right)
+
1.291
\right], 
\hspace{1.0cm} 
&&
P=\frac{\pi^{2}}{45}\frac{(kT)^{4}}{(\hbar c)^{3}}
+\frac{kT}{12V},
\label{sp1}
\end{eqnarray}
and equation (\ref{uc0}) is replaced by
\begin{equation}
U=\frac{\pi^{2}}{15}\frac{(kT)^{4}}{(\hbar c)^{3}} V+\frac{kT}{4}, \hspace{2.0cm} 
C_{V}=C+\frac{k}{4}.
\label{uc1}
\end{equation}
Therefore, $C_{V}$ is essentially the blackbody heat capacity at constant volume in equation (\ref{uc0})
up to a small correction.
Note also that the equation of state $PV=U/3$ still holds, but $C_{V}=3S$ holds only approximately, as a quick check shows.

An interesting fact is that $G=F+PV$ no longer vanishes identically. Instead,
by using equations (\ref{fss}) and (\ref{sp1}), it follows that
\begin{equation}
G=-\frac{kT}{4}\left[
{\rm ln}\left(\frac{kTV^{1/3}}{\hbar c}\right)
-
0.042
\right],	
\label{gss0}
\end{equation}
which, by noticing $P$ in equation (\ref{sp1}),
can be rewritten as a function of $T$ and $P$. Then, it is a simple matter
of calculation to show that equation (\ref{sv0}) leads correctly to the expressions
in equation (\ref{sp1}).

As $P$ now is a function of $T$ and $V$ [cf. equation (\ref{sp1})], equations
(\ref{c0}) and (\ref{akapa}) can be applied. Enthalpy $H=4U/3$, when expressed in
terms of $T$ and $P$, becomes
\begin{equation}
H=\frac{\pi^{2}}{135}\frac{(kT)^{5}}{(\hbar c)^{3}}
\left[P-\frac{\pi^{2}}{45}\frac{(kT)^{4}}{(\hbar c)^{3}}\right]^{-1}+\frac{kT}{3}.
\label{h0}
\end{equation}
Then equation (\ref{c0}) yields a well defined heat capacity at constant pressure, namely
\begin{equation}
C_{P}=\frac{4}{3}\frac{C^{2}}{k}
\left[1+\frac{75}{(4\pi)^{2}V}\left(\frac{\hbar c}{kT}\right)^{3} \right]+\frac{k}{3},
\label{c1}
\end{equation} 
where $C$ is the blackbody heat capacity at constant volume in equation (\ref{uc0}).
Clearly, up to small corrections, $C_{P}=4C^{2}/3k$.
Noticing equations (\ref{ht}) and (\ref{uc1}), one sees that $C_{P}/C_{V}\sim (kT/\hbar c)^{3}V$,
resulting that [see equation (\ref{uc1})]
\begin{equation}
C_{P}\gg C_{V}\approx C.
\label{c2}
\end{equation}
It is worth remarking that water volumetric heat capacity is 
$4.2 \times 10^{6} J/m^{3} K$, which corresponds to $10^{14}C$, when $V=1m^{3}$ and $T=300 K$,
and it is comparable to $10^{-2}C_{P}$ as can be promptly verified.

Recalling that $V$ can be written as a function of $T$ and $P$
by means of equation (\ref{sp1}),
rather lengthy  but straightforward calculations lead to [see equation (\ref{akapa})]
\begin{equation}
\alpha=\frac{16\pi^{2}}{15}k\frac{(kT)^{2}}{(\hbar c)^{3}}V+\frac{1}{T}, 
\hspace{2.0cm} 
\kappa_{T}=\frac{12 V}{kT},
\label{akapa1}
\end{equation}
which are also well defined quantities
\footnote{One may feel inclined to call $\kappa_{T}$ in equation (\ref{akapa1}) a
``classical'' quantity, since it does not contain $\hbar$. However, that is not appropriate.
The expression for $\kappa_{T}$ in equation (\ref{akapa1}) is just the $\hbar^{0}$-term of an expansion in powers of $\hbar$.}. At this point a nice test of consistency is available.
Namely, by noting equations (\ref{uc1}), (\ref{c1}) and (\ref{akapa1}), it follows that 
\begin{equation}
C_{P}-C_{V}=\frac{TV\alpha^{2}}{\kappa_{T}},
\label{identity1}
\end{equation}
which is a well known identity in thermodynamics \cite{cal85,hua87}.

Physical interpretation of the expressions in equation (\ref{akapa1}) is nearly self-evident in the light of the definitions in equation (\ref{akapa}). For example, as $\alpha$ in equation (\ref{akapa1})
is positive, 
it means that, keeping $P$ constant, the spherical shell expands as it warms.  
The positiveness of $\kappa_{T}$ in equation (\ref{akapa1}) will be addressed
later, in section \ref{ts}.

Before turning to the cubic box, it is worth calculating one more quantity in the spherical shell cavity, since it gives
another opportunity to check consistency of the results.  It happens to be the adiabatic compressibility $\kappa_{S}$, and it is defined as $\kappa_{T}$ in equation (\ref{akapa}), now 
replacing $T$ by $S$. One sets $dS=0$ in equation (\ref{sp1}) and writes $dT$ in terms of $dV$. 
Then, by considering $dP$ in equation (\ref{sp1}), it results that
\begin{equation}
\kappa_{S}=\frac{135}{4\pi^{2}}\frac{(\hbar c)^{3}}{(kT)^{4}}
\left[1+\frac{15}{4\pi^{2}V}\left(\frac{\hbar c}{kT}\right)^{3} \right]^{-1},
\label{kapas1}
\end{equation} 
where the factor out of the right brackets is $\kappa_{S}=3/4P$ from the usual 
blackbody thermodynamics. Now, by noticing equations (\ref{uc1}), (\ref{c1}), (\ref{akapa1}) and (\ref{kapas1}), it follows another thermodynamic identity, namely \cite{cal85,hua87}:
\begin{equation}
\frac{C_{V}}{C_{P}}=\frac{\kappa_{S}}{\kappa_{T}},
\label{identity2}
\end{equation}
as expected.

\section{Cubic box}
\label{cb}
Now, for a perfect conducting cubic box
carrying hot electromagnetic radiation, 
the procedures to obtain thermodynamics are as 
those followed 
in the previous section.
However, instead of the Helmholtz free energy given
in  equation (\ref{fss}),
one considers \cite{cas70,amb83}
(see appendix \ref{A})
\begin{equation}
F=-\frac{\pi^{2}}{45}\frac{(kT)^{4}}{(\hbar c)^{3}} V
+\frac{\pi}{4}\frac{(kT)^{2}}{\hbar c}V^{1/3},
\label{fcb}
\end{equation}
where smaller  corrections have been omitted again.
It should be stressed that,
due to the different geometries of the cavities,
equations (\ref{fss}) and (\ref{fcb}) differ only 
by their corrections to the familiar blackbody Helmholtz free energy [see equations (\ref{f1}) and (\ref{ht})]. Nevertheless, such a little difference will lead to contrasting results.

It follows from equation (\ref{fcb}) that [cf. equations (\ref{sp1}) and (\ref{uc1})]
\begin{eqnarray}
\hspace{-1.0cm} S=\frac{4\pi^{2}}{45}\left(\frac{kT}{\hbar c}\right)^{3}k V
-\frac{\pi}{2}\frac{k^{2}T}{\hbar c}\, V^{1/3}, 
\hspace{1.0cm} 
&&
P=\frac{\pi^{2}}{45}\frac{(kT)^{4}}{(\hbar c)^{3}}
-\frac{\pi}{12 V^{2/3}}\frac{(kT)^{2}}{\hbar c},
\label{sp2}
\end{eqnarray}
and
\begin{equation}
U=\frac{\pi^{2}}{15}\frac{(kT)^{4}}{(\hbar c)^{3}} V
-\frac{\pi}{4}\frac{(kT)^{2}}{\hbar c} V^{1/3}, \hspace{2.0cm} 
C_{V}=C-\frac{\pi}{2}\frac{k^{2}T}{\hbar c} V^{1/3}.
\label{uc2}
\end{equation}
Once more, $PV=U/3$ still holds, whereas $C_{V}=3S$ holds only approximately.
Here, it is worth pointing out that for a fixed $T$, as the volume $V$ of the cavity increases,
$P$ in equation (\ref{sp1}) decreases,
but $P$ in equation (\ref{sp2}) increases. It will be seen shortly that
this fact is relevant in the study of thermodynamic stability.

The Gibbs thermodynamic potential is also nonvanishing, i.e., 
\begin{equation}
G=\frac{\pi}{6}\frac{(kT)^{2}}{\hbar c}V^{1/3},	
\label{gcc0}
\end{equation}
and it differs considerably from that in equation (\ref{gss0})
for the spherical shell.
Again, it is a simple matter of calculation to verify that 
equations (\ref{sv0}) and (\ref{gcc0}) are consistent with
equation (\ref{sp2}). 

Corresponding to equation (\ref{h0}),
now one has for $H=4U/3$
\begin{equation}
H=\left(\frac{\pi}{3}\right)^{3/2}\frac{(kT)^{3}}{2(\hbar c)^{3/2}}
P\left[\frac{\pi^{2}}{45}\frac{(kT)^{4}}{(\hbar c)^{3}}-P\right]^{-3/2},
\nonumber
\end{equation}
and equation (\ref{c0}) yields
\begin{equation}
C_{P}=-\left(\frac{32C^{5}}{15\pi k^{2}}\right)^{1/3}
\left[1-\frac{45}{8\pi V^{2/3}}\left(\frac{\hbar c}{kT}\right)^{2} 
+\frac{225}{32\pi^{2} V^{4/3}}\left(\frac{\hbar c}{kT}\right)^{4} 
\right],
\label{c3}
\end{equation} 
which differs sharply from $C_{P}$ in equation (\ref{c1}),
especially because these heat capacities have different signs.
This feature will be addressed again in the context of 
thermodynamic stability.
Thus, up to small corrections, $C_{P}=-(32C^{5}/15\pi k^{2})^{1/3}$, and
noticing equation (\ref{uc2}) one sees that the appropriate version for 
equation (\ref{c2}) is now $|C_{P}|\gg C_{V}\approx C$.

The same steps that, in the previous section, have led to equations (\ref{akapa1}) and (\ref{kapas1}),
now lead to
\begin{equation}
\alpha=-\frac{8\pi}{5}\frac{k^{2}T}{(\hbar c)^{2}}V^{2/3}+\frac{3}{T}, 
\hspace{2.0cm} 
\kappa_{T}=-\frac{18}{\pi}\frac{\hbar c}{(kT)^{2}}V^{2/3},
\label{akapa2}
\end{equation}
and
\begin{equation}
\kappa_{S}=\frac{135}{4\pi^{2}}\frac{(\hbar c)^{3}}{(kT)^{4}}
\left[1-\frac{15}{4\pi V^{2/3}}\left(\frac{\hbar c}{kT}\right)^{2} \right]^{-1}.
\nonumber
\end{equation} 
Noteworthy is the fact that $\alpha$ and $\kappa_{T}$ in equation (\ref{akapa2}) are negative,
whereas their counterparts for the spherical shell in section {\ref{ss}} are positive [cf. equation (\ref{akapa1})]. Regarding $\alpha$, equations (\ref{akapa}) and (\ref{akapa2})  say that, keeping $P$ constant, the cubic box shrinks as it warms, instead of expanding as the spherical shell.  
The negativeness of $\kappa_{T}$ in equation (\ref{akapa2}) will be considered shortly.

Therefore, $\alpha$, $C_{P}$ and $\kappa_{T}$ are also well defined quantities in 
the cubic box and, together with $C_{V}$ and $\kappa_{S}$, they satisfy the usual identities in equations (\ref{identity1}) and (\ref{identity2}), as should be.

\section{Thermodynamic stability}
\label{ts}
A long-lasting physical system must satisfy both the following criteria 
for stable thermodynamic equilibrium \cite{cal85}:
\begin{equation}
C_{V}>0, \hspace{2.0cm}
\kappa_{T}>0.
\label{ste}
\end{equation}
The first inequality in equation (\ref{ste}) implies thermal stability, whereas the second
one implies mechanical stability.
By examining the results in the previous sections, one sees that blackbody radiation is
a stable thermodynamic system in the spherical shell; but it is not in 
the cubic box, since the isothermal compressibility in the latter is negative [see equation (\ref{akapa2})]. Note that, for both cavities, $C_{V}>0$ and $\kappa_{S}>0$,
thus the negativiness of $C_{P}$ in equation (\ref{c3}) agrees with equations (\ref{identity1}) and (\ref{identity2}), as expected.  

It should be remarked that, by taking into account the definition of $\kappa_{T}$ in equation
(\ref{akapa}), another way of expressing the criterion for mechanical stability is $(\partial_{V}P)_{T}<0$, i.e., pressure must diminish as the system volume increases, with the temperature kept fixed. 
Although thermodynamic stability requires $\kappa_{T}>0$, there exist indeed substances with negative
isothermal compressibities \cite{bau98}. Most likely they have to be monitored closely
to prevent ``collapse''.

\section{Final remarks}
\label{fr}
In ordinary blackbody thermodynamics, it is assumed that radiation pressure $P$ on the walls of the cavity depends on temperature $T$ alone. That is only part of the story, though, since sub-leading contributions in the expression for $P$ carries also dependence on volume $V$. This fact then allows definition of the quantities $C_{P}$, $\alpha$ and $\kappa_{T}$, as this work has shown for two perfect conducting cavities, namely, a spherical shell and a cubic box.

It is perhaps worth highlighting that, in leading order, $C_{P}=4C^{2}/3k$ for the spherical shell, whereas for the cubic box $C_{P}=-(32C^{5}/15\pi k^{2})^{1/3}$,
with $C$ denoting the familiar blackbody heat capacity at constant volume. These radically different expressions for $C_{P}$ mirror the fact that statistical mechanics of the normal modes in the cavity are affected by the cavity's geometry. By using well known criteria,
it was shown that blackbody radiation in the spherical shell is thermodynamic stable,
while in the cubic box is not, and that is related with the negative $C_{P}$ above.

The fact that  blackbody radiation in a perfect conducting spherical shell is a stable thermodynamic system, unlike in a perfect conducting cubic box, is indeed puzzling, especially considering that the regime studied here is that of high temperatures and/or large cavities. It comes to mind at which point the cavities geometry becomes relevant on this matter, by looking at cavities with various geometries.

Since $\alpha$, $\kappa_{T}$ and $\kappa_{S}$ are quantities amenable to experimental determination, one may wonder whether the findings in this work can be experimentally validated. 
Perhaps a good start is the adaptation of \cite{kub23} for a blackbody cavity.
It consists of a new method of determining $\kappa_{T}$ of molecular systems by simulation where 
average pressures and volumes play a role in the very definition of $\kappa_{T}$ [see equation (\ref{akapa})]. Another new ingredient in the method is that
the external equilibrium pressure is kept by a spring attached to the cavity's wall.
It is conceivable that for a macroscopic blackbody cavity, the simulation data in \cite{kub23}
can be replaced by experimental data to determine $\kappa_{T}$ in the lab.
It is worth recalling that from $\kappa_{T}$ one can get other quantities
[see, e.g., equations (\ref{identity1}) and (\ref{identity2})].

A last remark is in order. Here high temperatures and/or large cavities have been addressed
[see equation (\ref{ht})]. It would be pertinent to extend the present study to determine
$C_{P}$, $\alpha$ and $\kappa_{T}$ when $T$ is low and/or $V$ is small. Such an investigation has been undertaken in \cite{mor23} for the geometry
of a slab, revealing unusual and interesting features due to the strong presence of vacuum polarization.

\appendix
\section{Derivation of equations (\ref{fss}) and (\ref{fcb})}
\label{A}
In this appendix, for convenience, fundamental constants have been set equal to unity.
As is well known, statistical mechanics of hot radiation at temperature $1/\beta$,
in a cavity of volume $V$,
begins with the calculation of the Helmholtz free energy \cite{hua87}, i.e.,
\begin{equation}
F=U_{0}+\frac{1}{\beta}\sum_{\bf k}
\ln\left(1-e^{-\beta\omega_{{\bf k}}}\right),
\label{F}
\end{equation}
where the vacuum energy $U_{0}$ will be neglected, since one is interested here in the regime of high temperatures, when $\beta\rightarrow 0$. If the geometry and boundary conditions on the cavity's walls are
both simple, then the summations over the quantum numbers ${\bf k}$ in equation (\ref{F}) can be transformed into integrations plus corrections by means of the Abel-Plana formula, for example.
However, in dealing with a spherical shell and perfect conducting walls, a more effective approach needs to be used, such as the ``$\zeta$-function regularization''
\footnote{See, e.g., the excellent review \cite{nes04}.}. 
At this point, it should be recalled that $\omega_{{\bf k}}^{2}$ in equation (\ref{F})
are eigenvalues of minus the Laplacian operator, $-\Delta$, corresponding to the particular geometry of the cavity and boundary conditions on its walls. 

Without going too much into technicalities,  $\zeta$-function regularization
yields the following high temperature expansion for the the Helmholtz free energy 
[cf. equation (\ref{f1})]
\begin{equation}
F=-\frac{\pi^{2} V}{45\beta^{4}}
-\frac{B_{1/2}\, \zeta_{R}(3)}{4\pi^{3/2}\beta^{3}}
-\frac{B_{1}}{24\beta^{2}}
+\frac{B_{3/2}\, \ln(\beta)}{(4\pi)^{3/2}\beta}
-\frac{\zeta'(0)}{2\beta}+\cdots,
\label{f3}
\end{equation}
which holds when $\beta\rightarrow 0$, and where $\zeta_{R}$ is the familiar Riemann zeta-function,
$B_{1/2}$, $B_{1}$, $B_{3/2}$
are heat kernel coefficients, and $\zeta'(0)$ is the derivative of the spectral zeta-function at zero argument.

For a perfect conducting spherical shell of radius $a$, the relevant nonvanishing heat kernel coefficient in equation (\ref{f3}) is $B_{3/2}=2\pi^{3/2}$, and  $\zeta'(0)=0.38429+\ln(\sqrt{a})$. Thus, equation (\ref{f3}) yields equation (\ref{fss}).

In the case of a perfect conducting rectangular box of sides $a$, $b$ and $c$, 
now the relevant nonvanishing heat kernel coefficient is $B_{1}=-2\pi(a+b+c)$.
Again, equation (\ref{f3}) yields equation (\ref{fcb}).
Perhaps it should be stressed that $B_{1/2}$ in equation (\ref{f3}) vanishes for both perfect conducting cavities.

\vspace{1cm}
\noindent{\bf Acknowledgements} -- 
Work partially supported by
``Funda\c{c}\~{a}o de Amparo \`{a} Pesquisa do Estado de Minas Gerais'' (FAPEMIG)
and by ``Coordena\c{c}\~{a}o de Aperfei\c{c}oamento de Pessoal de N\'{\i}vel Superior'' (CAPES).


\begin{thebibliography}{88}


\bibitem{pat72} 
R. K. Pathria,
\emph{Statistical Mechanics},
Elsevier, UK (1972).


\bibitem{cal85} 
H. B. Callen, 
\emph{Thermodynamics and an Introduction to Thermostatistics},
John Wiley \& Sons, U.S.A. (1985).


\bibitem{hua87}
K. Huang, 
\emph{Statistical Mechanics}, 
John Wiley \& Sons, U.S.A. (1987).



\bibitem{kel81}
R. E. Kelly, Thermodynamics of blackbody radiation,
Am. J. Phys.
{\bf 49}, 714 (1981).


\bibitem{rod89}
R. F. Rodr\'{\i}guez and L. S. Garc\'{\i}a-Col\'{\i}n,
The specific heat puzzle in black-body radiation,
Eur. J. Phys. 
{\bf 10}, 
214 (1989).




\bibitem{dav82} 
N. D. Birrel and P. C. W. Davies,
\emph{Quantum Fields in Curved Space}
(Cambridge University Press, Cambridge UK, 1982)


\bibitem{mil01}
K. A. Milton, 
\emph{The Casimir Effect, Physical Manifestations
of Zero-Point Energy}, 
World Scientific, New Jersey (2001).


\bibitem{bor01}
M. Bordag, U. Mohideen and V. M. Mostepanenko,
New developments in the Casimir effect,
Phys. Rept.
{\bf 353}, 1 (2001).



\bibitem{cas48}
H. B. G. Casimir, 
On the attraction between two perfectly conducting plates,
Proc. K. Ned. Akad. Wet. {\bf 51}, 793 (1948).


\bibitem{boy68}
T. H. Boyer, 
Quantum electromagnetic zero-point energy of a conducting spherical shell
and the Casimir model for a charged particle, 
Phys. Rev.
{\bf 174}, 1764
(1968). 



\bibitem{lif56}
E. M. Lifshitz, 
The Theory of Molecular Attractive Forces between Solids,
Zh. Eksp. Teor. Fiz. {\bf 29}, 94 (1955) [Sov. Phys.
JETP {\bf 2}, 73 (1956)].

\bibitem{meh67}
J. Mehra, 
Temperature correction to the Casimir effect,
Physica {\bf 37}, 145 (1967).


\bibitem{bro69}
L. S. Brown and G. J. Maclay, 
Vacuum Stress between Conducting Plates: An Image Solution, 
Phys. Rev. {\bf 184}, 1272 (1969).


\bibitem{cas70}
K. M. Case and S. C. Chiu,
Electromagnetic Fluctuations in a Cavity,
Phys. Rev. {\bf A 1}, 1170 (1970). 




\bibitem{bal78}
R. Balian and B. Duplantier, 
Electromagnetic waves near perfect conductors. II. Casimir effect,
Ann. Phys.  {\bf 112}, 165
(1978)


\bibitem{sch78}
J. Schwinger, L. L. DeRaad, Jr., and K. A. Milton,
Casimir Effect in Dielectrics,
Ann. Phys.  {\bf 115}, 1
(1978).


\bibitem{dow78}
J. S. Dowker and G. Kennedy,  
Finite temperature and boundary effects in static space-times, 
J. Phys. {\bf A 11}, 895 
(1978).



\bibitem{amb83}
J. Ambj\o rn and S. Wolfram, 
Properties of the Vacuum. I. Mechanical and Thermodynamic, 
Ann. Phys.  {\bf 147}, 1 (1983).

\bibitem{bau98}
R. H. Baughman, S. Stafstr\"{o}m, C. Cui and S. O. Dantas,
Materials with Negative Compressibilities in One or More Dimensions,
Science {\bf 279}, 1522 (1998).



\bibitem{kub23}
K. Kubota, A. Firoozabadi and L. F. M. Franco,
A new method to compute the isothermal
compressibility of confined fluids,
SSRN,
https://ssrn.com/abstract=4341625.




\bibitem{mor23}
E. S. Moreira Jr. and  H. da Silva,
Blackbody thermodynamics in the presence of Casimir's effect,
J. Stat. Mech. 063102 (2023).



\bibitem{nes04}
V. V. Nesterenko, G. Lambiase and G. Scarpetta, 
Calculation of the Casimir energy at zero and finite temperature: Some recent results,
Riv. Nuovo Cim. {\bf 27}, 1 (2004). 































\end{thebibliography}
\end{document}